**Phonon and electronic properties of semiconducting silicon nitride bilayers**

Jiesen Li [1,#], Wanxing Lin [2,3,#], Junjun Shi[1], Feng Zhu[1], Haiwen Xie[1], Dao-Xin Yao [2,4 †],

[1] School of Environment and Chemical Engineering, Foshan University, Foshan 528000, P. R. China

[2] State Key Laboratory of Optoelectronic Materials and Technologies, Guangdong Provincial Key Laboratory of Magnetoelectric Physics and Devices, School of Physics, Sun Yat-Sen University, Guangzhou 510275, P. R. China

[3] School of Materials Science and Engineering, Guangdong Ocean University, Yangjiang 529500, P. R. China.

[4] International Quantum Academy, Shenzhen 518048, P. R. China.

[#] Two authors contribute equally.

[†] Corresponding author. Tel: 86 020 84112078. E-mail: yaodaox@mail.sysu.edu.cn (Dao-Xin Yao)

**Abstract**

The two-dimensional (2D) IV-V semiconductors have attracted much attention due to their fascinating electronic and optical properties. In this work, we predicted three phases of silicon nitrides, denoted $\alpha$-$Si_2N_2$, $\beta$-$Si_2N_2$, and $\gamma$-$Si_4N_4$, respectively. Both $\alpha$-$Si_2N_2$ and $\beta$-$Si_2N_2$ consist of two buckled SiN sheets, and $\gamma$-$Si_4N_4$ consists of two puckered SiN sheets. It is challenging to transform between $\alpha$-$Si_2N_2$ and $\beta$-$Si_2N_2$ because of the high energy barrier. The three dynamically stable bilayers are semiconductors with fundamental indirect band gaps from 0.25 eV to 2.92 eV. As expected, only the s and p orbitals contribute to the electronic states, and the $p_z$ orbitals dominate near the Fermi level. Furthermore, insulator-metal transitions occur in $\alpha$-$Si_2N_2$ and $\beta$-$Si_2N_2$ under the biaxial strain of 16%. These materials perhaps have potential applications in microelectronics and spintronics.

## 1. Introduction

2D materials have attracted significant attention since the discovery of graphene [1]. Studies on graphene-like monolayers of group IV elements is a hot topic in condensed matter and material science. Silicene and germanene were proposed early [2] and later confirmed experimentally [3-5]. The monolayer stanene made up of tin atoms was successfully synthesized by epitaxial growth [6]. Besides, the plumbene was predicted and found to be a topological insulator [7, 8]. In addition to the group IV 2D materials, pnictogen monolayers are also extensively investigated. The black phosphorus with high mobility was exfoliated [9-11], and the blue phosphorus with a tunable gap was proposed [12, 13] and also realized in the experiment [14, 15]. It was found that the free-standing monolayers of phosphorus are theoretically possible [16, 17]. Furthermore, the honeycombed nitrogene was also proposed, which presents a Dirac cone under the suitable biaxial strain [18, 19]. The binary compounds of pnictogen were also under intensive studies [20, 21].

The monolayers of binary compounds, including group IV and group V elements, are recent emerging 2D materials. The semiconducting 2D SiP materials have potential applications in photohydrolytic catalysts due to their novel optical properties [22]. The graphene-like IV−V narrow-gap semiconductors, such as $Si_3N$ and $Si_3P$, show significant anisotropy, and with a narrow gap, and demonstrate excellent optical absorption [23]. The silicon-based IV−V compounds show excellent photocatalysts in water-splitting, and the graphene/SiAs heterojunction is a potential candidate in nanoelectronic devices [24]. Remarkably, the silicon nitride harbors several known phases of three-dimensional (3D) structure, such as $\alpha$-$Si_3N_4$, $\beta$-$Si_3N_4$, $\gamma$-$Si_3N_4$, c-$Si_3N_4$, *etc*. These $Si_3N_4$ exhibit outstanding hardness and thermal stabilities, also extensively applied in the ceramic industry [25-27]. While the IV-V bilayers and their band structures dependence on the biaxial strain were less investigated.

As coinstantaneous research of our previous works [28, 29], this work proposed three phases of IV-V compounds, denoted $\alpha$-$Si_2N_2$, $\beta$-$Si_2N_2$, and $\gamma$-$Si_4N_4$ bilayers, respectively. The phonon dispersions and electronic properties were systematically investigated using the density functional theory (DFT). While it was found that both $\alpha$-$Si_2N_2$ and $\beta$-$Si_2N_2$ bilayers are thermodynamically and thermodynamically stable, and the $\gamma$-$Si_4N_4$ may be thermodynamically stabilized by suitable substrates. All bilayers show a semiconducting characteristic in the free-standing state, and the α-$Si_2N_2$ and β-$Si_2N_2$ undergo insulator-metal transitions under tensile strain. Furthermore, the electrons in the SiN monolayer are spin-polarized, and the sub-lattice presents a magnetic moment of $1\ \mu_B$. These proposed materials may be synthesized from silica with ammonia in the gas phase, and may have potential applications in spintronics and straintronics.

## 2. Calculation details

The calculations were based on the density functional theory (DFT), as implemented in the Vienna *Ab initio* Simulation Package (VASP) code [30]. The plane augmented wave (PAW) with a cutoff energy of 500 eV was used as the basis, and Perdew-Burke-Ernzerh (PBE) functional [31] was treated as the exchange-correlation. Heyd-Scuseria-Ernzerhof hybrid functional (HSE06), which is based on a screened Coulomb potential [32, 33], was also used for electronic structure calculation to compare with the PBE results. The systems were restricted to periodic boundary conditions along with the sheets. Neighboring bilayers were separated by a vacuum of at least 15 Å thick to eliminate the interlayer couplings. Ions were relaxed until the net force on each ion was less than $10^{-5}$ eV/Å. The Brillouin zone (BZ) was sampled by a 15×15×1 Γ-centered grid. Bader charge analysis is based on an improved grid-based algorithm implemented in the bader code [34]. Phonon dispersions calculation was performed on the VASP interface of Phonopy [35], and the force constants between atoms were calculated on a 4×4×1 supercell. The reaction path was calculated by the climbing image nudged elastic band method (CINEB) [36-38], with three images being inserted between the structures of two phases. Then the transition state obtained by CINEB was further verified by improved dimer techniques [39, 40], and the vibration frequencies were calculated by diagonalization of Hessian matrixes. The spin-orbit coupling (SOC) was not included in all calculations because it is negligibly small. The electronic properties under strain were investigated based on the PBE calculation.

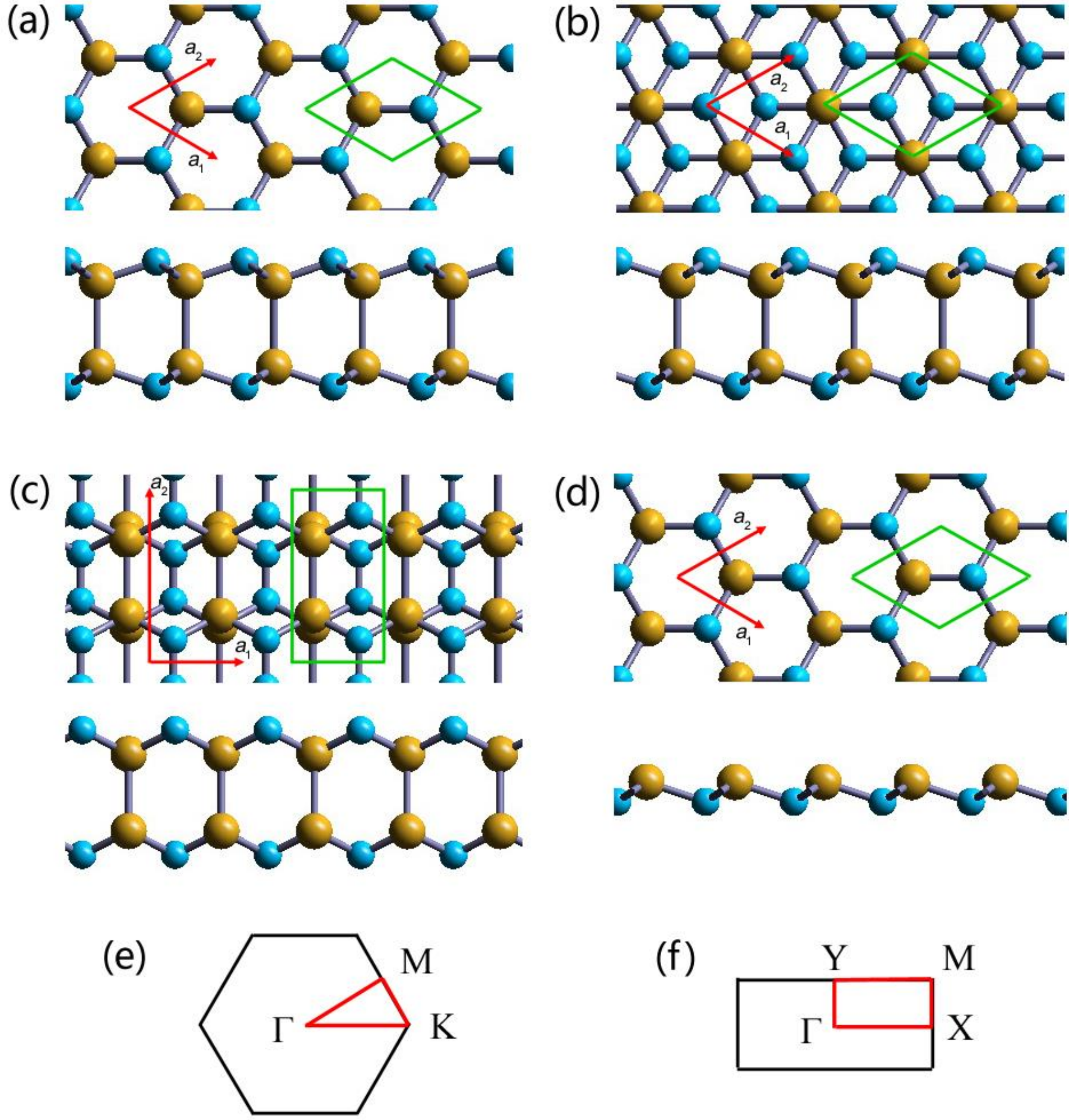

Figure 1. The top and side view of the crystal structures of (a) α-$Si_2N_2$, (b) β-$Si_2N_2$, (c) γ-$Si_4N_4$ bilayers, and (d) SiN monolayer. The orange and light blue spheres denote silicon and nitrogen atoms, respectively. The red arrows indicate the basis vectors, and the areas enclosed by the green lines denote the unit cells. The first BZ and high symmetry points of (e) α-$Si_2N_2$, β-$Si_2N_2$, SiN monolayer, and (f) γ-$Si_4N_4$.

## 3. Result and discussion

### 3.1 Structures and stabilities

The crystal structures of α-$Si_2N_2$ and β-$Si_2N_2$ can be considered as the AA stacking and AB stacking of the buckled SiN monolayers, respectively. The γ-$Si_4N_4$ also consists of two puckered SiN monolayers, while the structure is slightly different, as shown in Fig. 1. For the α-$Si_2N_2$ and β-$Si_2N_2$, the relaxed lattice constant is 2.88 Å, the Si-Si bond is 2.42 Å, and the Si-N bond is 1.76 Å. The Si-Si bond, corresponding to the interlayer distance, is stretched from the usual 2.22 Å to 2.42 Å. The bond angle between the two Si-N bonds is 110º24’, and the one between the Si-N bond and the Si-Si bond is 108º30’. Both slightly differ from the regular tetrahedron 109º28’, caused by the nonequivalent hybridization on the central carbon atom that arises from the repulsion between the lone pairs of nitrogen. The β-$Si_2N_2$ is only 59 meV for each unit cell higher than α-$Si_2N_2$ in energy. For γ-$Si_4N_4$, the N-N bond is 1.47 Å, the intralayer Si-Si bond is 2.34 Å, and the Si-N bond is 1.78 Å. The interlayer Si-Si bond is 2.40 Å, corresponding to the interlayer distance.

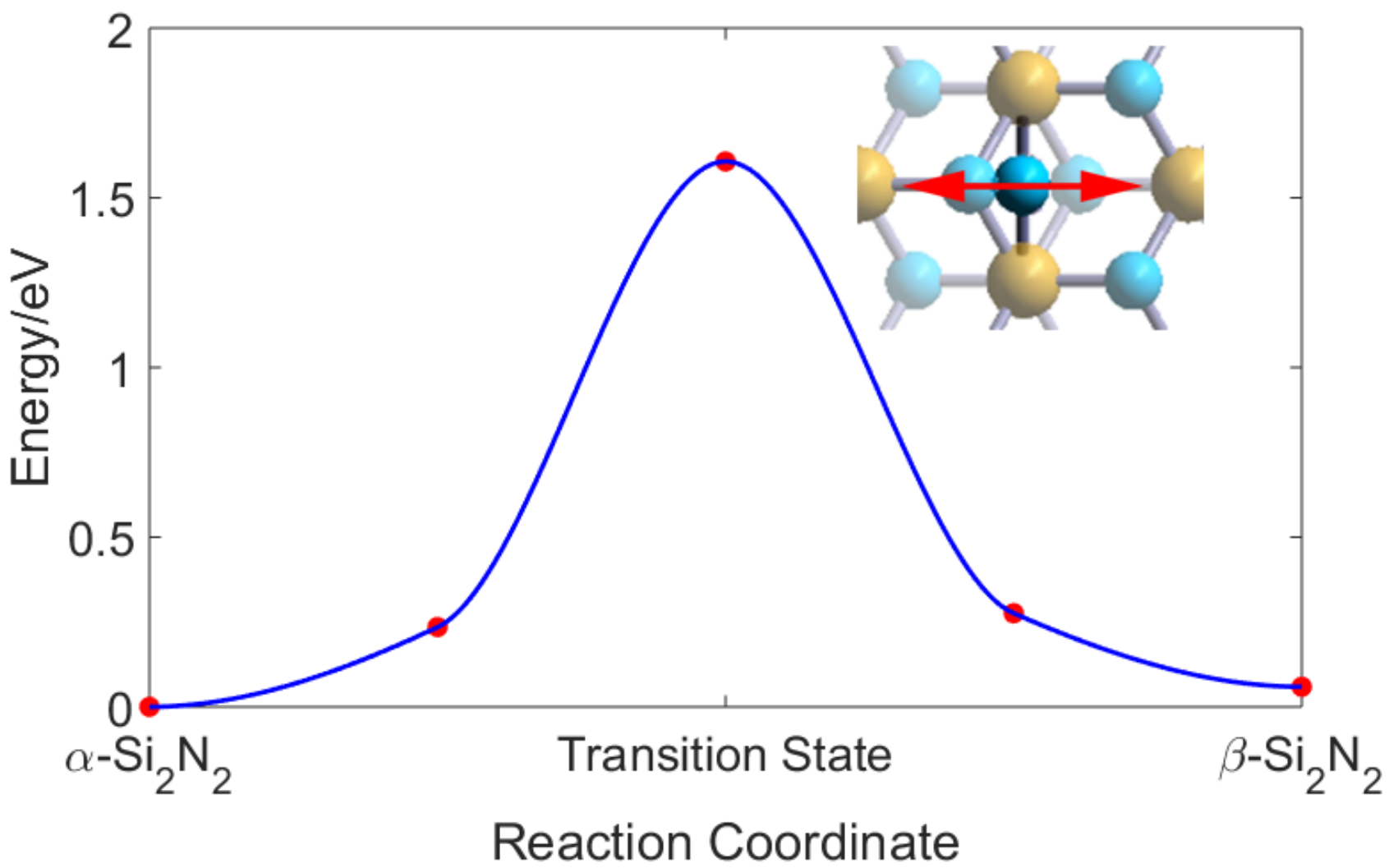

Figure 2. The evolution of total energy during the transformation from α-$Si_2N_2$ to β-$Si_2N_2$. The energy of α-$Si_2N_2$ was set to zero. Red dots denote the inserted images during CINEB calculation; the first, the third, and the last dot correspond to α-$Si_2N_2$, transition state, and β-$Si_2N_2$, respectively. The blue curve is the spline interpolation. The transition states during the transformation are shown in the inset. Red double-ended arrow indicates the vibration mode with imaginary frequency, which corresponds to the conversion path in the transition state.

The transition between α-$Si_2N_2$ and β-$Si_2N_2$ can be achieved by breaking and reforming the Si-N bonds, as illustrated in Fig. 2. In both bilayers, one nitrogen atom is bonded to three silicon atoms. In the transition state, one of the three Si-N bonds is broken. The nitrogen atom has an imaginary frequency of 11.3 THz in the transition state, and the corresponding vibration mode is indicated as the double-ended arrow in the inset of Fig. 2. As mentioned above, the α-$Si_2N_2$ is 59 meV lower than the β-$Si_2N_2$ in energy for each unit cell, that the activation energy is slightly different for forward and backward transformation; the system has to overcome 1.61 eV to transform from α-$Si_2N_2$ to β-$Si_2N_2$, or 1.55 eV from β-$Si_2N_2$ to α-$Si_2N_2$. In other words, the conversion from α-$Si_2N_2$ to β-$Si_2N_2$ is relatively difficult due to the high activation barrier, because it involves bond breaking.

TABLE 1. The wave velocities of acoustic modes (m/s)

| Acoustic modes | α-$Si_2N_2$ | | β-$Si_2N_2$ | |
|---|---|---|---|---|
| | Γ→M | Γ→K | Γ→M | Γ→K |
| ZA | 1256 | 1249 | 1183 | 1092 |
| TA | 6308 | 6384 | 7672 | 7829 |
| LA | 13986 | 13917 | 14590 | 14388 |

The phonon dispersions were calculated to test the stabilities of the silicon nitrides, as shown in Fig. 3. No vibration mode with imaginary frequency appears in the whole BZ for α-$Si_2N_2$ and β-$Si_2N_2$, suggesting both structures are dynamically stable. Additionally, the phonon dispersions show that both phases are mostly isotropic, and the wave velocities of the acoustic modes are listed in Table 1. While the differences in wave velocities indicate β-$Si_2N_2$ is slightly more anisotropic than α-$Si_2N_2$. These velocities, which reflects the rigidity, are comparable to graphene [41, 42]. it reveals that α-$Si_2N_2$ and β-$Si_2N_2$ are more rigid than most other 2D materials. However, for the γ-$Si_4N_4$, vibration modes with slight imaginary frequencies appear along the Γ-Y line in the BZ. These imaginary frequencies may be caused by the softening of phonon modes, indicating that it could be stable under certain conditions, for example, it may be stabilized by proper substrate. The lone pairs on nitrogen atoms may be responsible for this softening of the phonon mode of γ-$Si_4N_4$, because, instead of pointing perpendicular to the plane, lone pairs in the $sp^3$ hybridized orbitals on the two neighboring nitrogen atoms are slightly tilted towards each other, and therefore increase the electrostatic repulsion. The projected density of states (PDOS) of phonon reveals that the optical modes with high frequencies are mainly contributed by the nitrogen atoms, and both elements have approximately equal contribution to the low and medium modes. In addition, *ab-initio* molecular dynamic simulations show that these three structures are thermodynamically stable, and the results are available in Fig. S(1-3) of the Supplementary Material. On the other hand, the pronounced imaginary frequencies indicate that the SiN monolayer is unstable in the free-standing state.

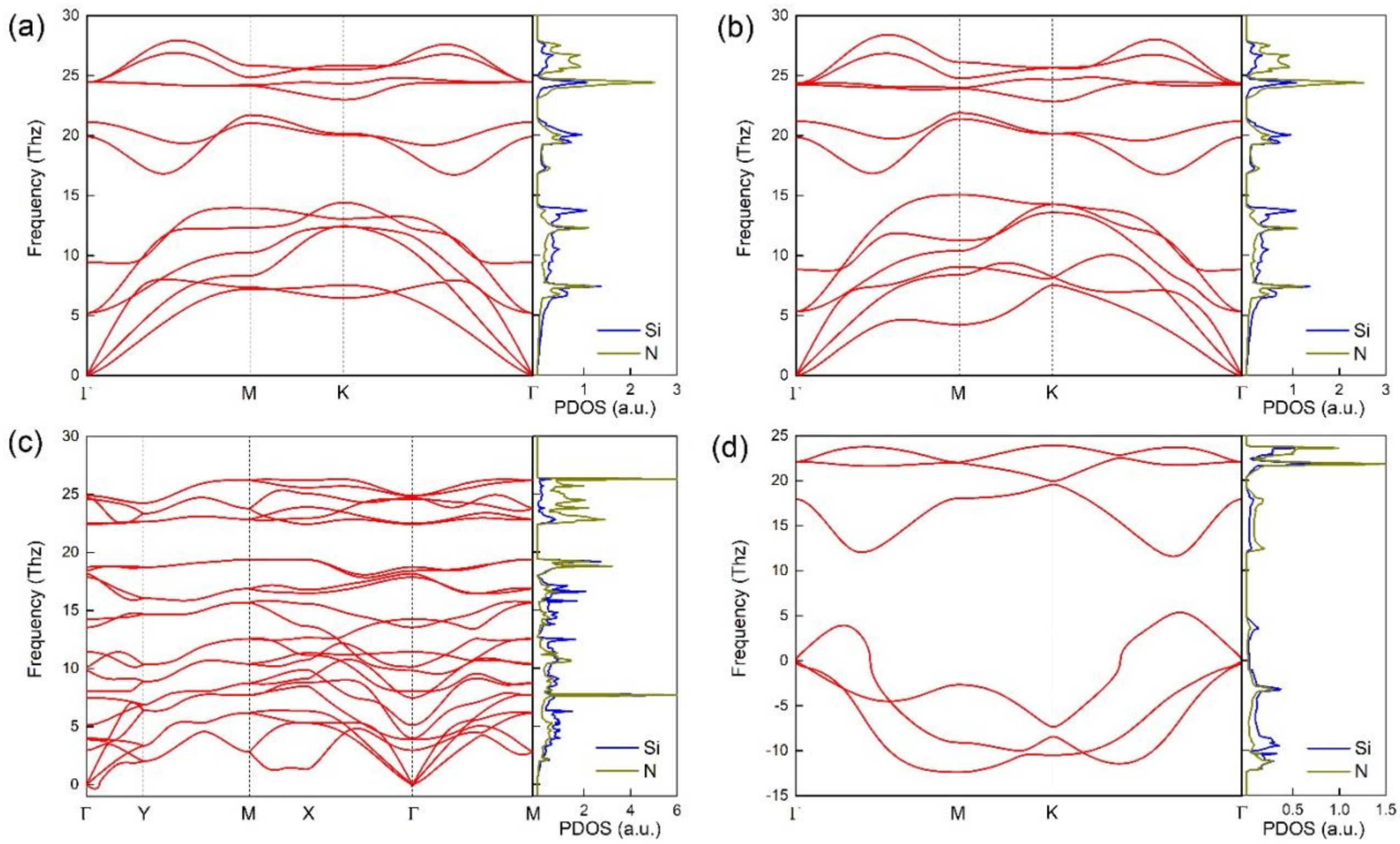

Figure 3. The phonon dispersions of (a) $\alpha$-$Si_2N_2$, (b) $\beta$-$Si_2N_2$, (c) $\gamma$-$Si_4N_4$, and (d) SiN monolayer. The left and right parts of each panel denote the phonon band and the PDOS on each atom, respectively.

A $4 \times 4 \times 1$ supercell was relaxed to verify the instability of the SiN monolayer. As shown in Fig. 4(a-c), wraps and ripples are formed in the monolayer of SiN. Some bonds are twisted and turned, and some bonds are even broken or reformed, indicating their instability. This may be caused by the dangling bonds on Si atoms. On the other hand, analysis of vibration modes in $\gamma$-$Si_4N_4$ with imaginary frequencies also shows that the lone pairs on nitrogen atoms repel each other since these vibrations point along the Y-axis, as shown in Fig. 4(d). As mentioned above, the repulsion of the lone pairs may lead to the softening of phonon modes near the $\Gamma$ point, and thus lead to imaginary frequencies.

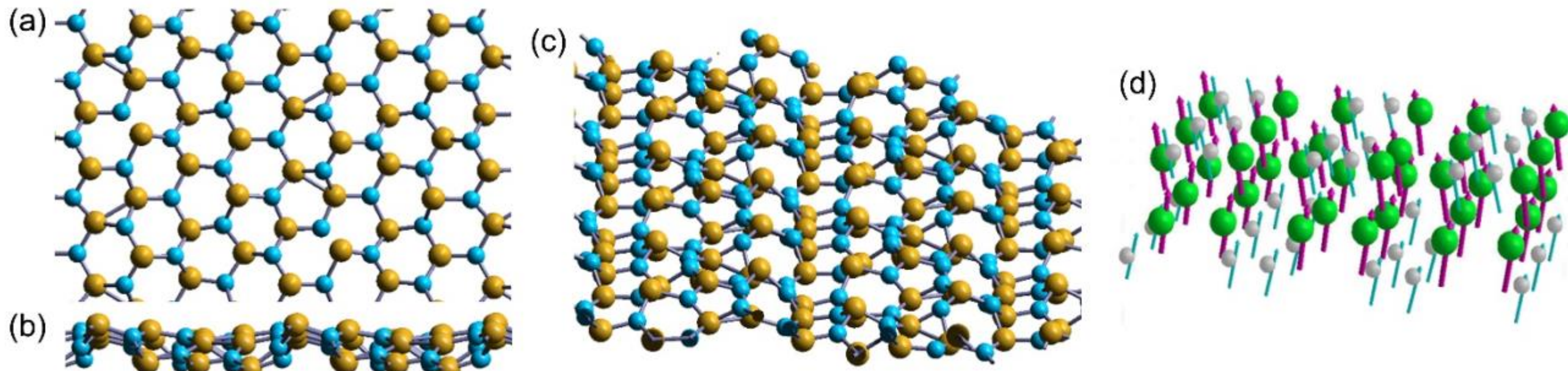

Figure 4. (a) Top view, (b) side view, and (c) bird's-eye view of fully relaxed $4 \times 4 \times 1$ supercell of SiN monolayer. The orange and light blue spheres denote silicon and nitrogen atoms, respectively. (d) One of the vibration modes with imaginary frequency for γ-$Si_4N_4$. Green and grey spheres denote silicon and nitrogen atoms, respectively. The arrows represent the vibration directions.

3.2 Electronic properties

Compared with nitrogene, the band structures of α-$Si_2N_2$ and β-$Si_2N_2$, as shown in Fig. 5 (a and b), have smaller indirect band gaps of 1.74 eV and 1.77 eV based on the PBE calculation, respectively [18]. On the HSE06 level, the band gaps become 2.75 eV and 2.92 eV, respectively. For both cases, the conducting band minimum (CBM) locates at the M point in the BZ. The conducting band of α-$Si_2N_2$ in K is very close to CBM in energy. For α- and β-$Si_2N_2$, the valance band maximum (VBM) deviates from the Γ points. Based on the analysis of density of states (DOS), which is available in Fig. S(4-5) of the Supplementary Material, most states of α-$Si_2N_2$ and β-$Si_2N_2$ below the Fermi level come from the nitrogen atoms, while a more significant portion of states above the Fermi level originate from silicon atoms. This can be explained in terms of electronegativity; as nitrogen is more electronegative than silicon, the orbitals in the outermost shell have lower energies in nitrogen than in silicon. There is a clear separation of s and p orbitals in energy. For nitrogen, the s orbital appears almost exclusively in the innermost branch, even though there are a small amount of s characters near the Fermi level. In the energy windows of -10 eV to -3 eV, the bands have a robust $p_{x+y}$ character, but the $p_z$ orbital dominates near the Fermi level. The PBE calculation indicates the γ-$Si_4N_4$ is metallic, while the HSE06 result suggests it is an indirect insulator with a band gap of 0.25 eV. For γ-$Si_4N_4$, the CBM is at X point, and the VBM is along the X-Γ line, as shown in Fig. 5(c).

The band structure indicates SiN monolayer is also a narrow-gap semiconductor. The band structure of the SiN monolayer on the PBE level is shown in Fig. 5 (d). Electrons in the SiN monolayer are spin-polarized, and the occupied states are closely related to the dangling bonds on Si atoms and the lone pairs on N atoms. The $p_z$ orbital of both types of atoms has major contribution to the states near the Fermi level. More details are available in Fig. S7 of the Supplementary Material. The dangling bonds on silicon atoms are mainly responsible for the instability of SiN monolayer, as well as a sub-lattice magnetization of $1\ \mu_B$.

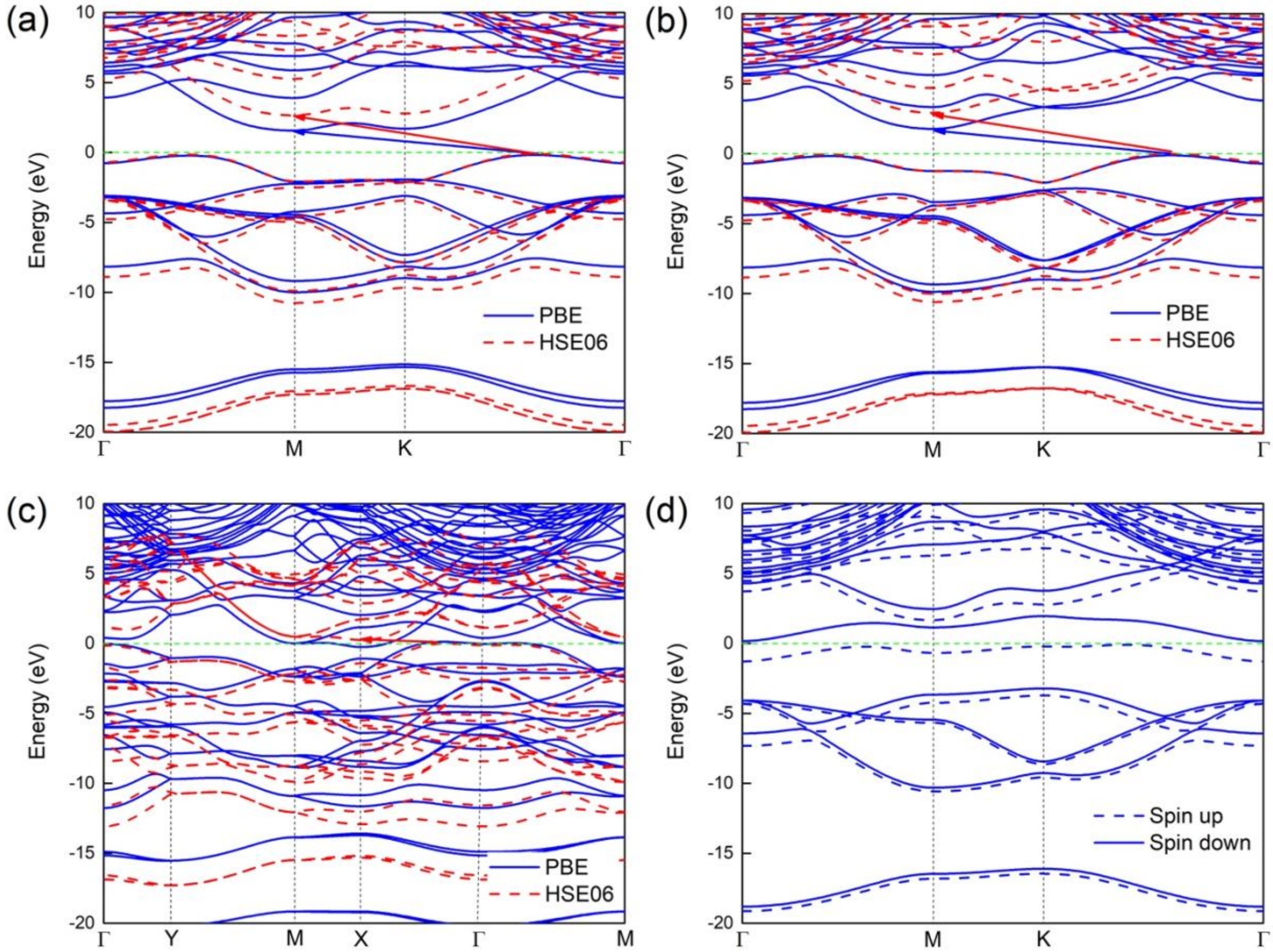

Figure 5. The band structures of (a) α-$Si_2N_2$, (b) β-$Si_2N_2$, (c) γ-$Si_4N_4$, and (d) SiN monolayer. In (a-c), the solid blue lines refer to band structures on the PBE level, while the red dash ones refer to the band structure on the HSE06 level. The blue and red arrows indicate the directions from VBM to CBM based on the PBE and HSE calculations, respectively. In (d), the blue dash and solid blue curves correspond to spin-up and spin-down components of the bands on the PBE level. The Fermi energy is set to zero.

To analyze the bonding of the materials, the band-decomposed charge distributions in the energy window of -2 eV to 0 eV with respect to Fermi energy were calculated based on the PBE calculation.

For $\alpha$-$Si_2N_2$, $\beta$-$Si_2N_2$, and $\gamma$-$Si_4N_4$, the band-decomposed charge is mainly distributed around the nitrogen atoms and the bonding area between nitrogen atoms and silicon atoms, as shown in Fig. 6 (a-c). Based on the PDOS in Fig. S(4-6) of the Supplementary Material, the states near the Fermi energy are mainly contributed by the $p_z$ orbitals, and the $p_z$ orbitals of the two neighboring silicon atoms forming a Si-Si bond, which is consistent with the distribution of band-decomposed charges. Both PBE and HSE06 calculated PDOS of the SiN monolayer are available in Fig. S7 of the Supplementary Material. For the spin-polarized band-decomposed charge distribution of the SiN monolayer, the band-decomposed charge distribution around the nitrogen atoms is roughly the same with the $\alpha$-$Si_2N_2$ and $\beta$-$Si_2N_2$, but the distribution around the silicon atoms is similar to the distribution of the $sp^3$ hybridized orbitals, as shown in Fig. 6(d). The overlap of these $sp^3$ hybridized orbitals forms the bonds in $\alpha$-$Si_2N_2$ and $\beta$-$Si_2N_2$; the case of $\gamma$-$Si_4N_4$ is similar. Since N is more electronegative than Si, the Si-N bond is polar, with N carrying more negative charge. Bader charge analysis shows that, 1.41 electrons transfer from Si to N in $\gamma$-$Si_4N_4$. In contrast, 2.21 electrons transfer from Si to N in $\alpha$- and $\beta$-$Si_2N_2$, which means the electron transfer is more significant and forms stronger bonds in $\alpha$- and $\beta$-$Si_2N_2$ compared with $\gamma$-$Si_4N_4$.

The semiconductors with intrinsic wide gaps have an extensive application in photocatalysis and optoelectronics due to their tunable intrinsic band gaps. The band gaps of $\alpha$-$Si_2N$ and $\beta$-$Si_2N_2$ can be tuned using the biaxial tensile strain. The strain may be induced by lattice mismatch, and can be further controlled by temperature due to a mismatch of the thermal coefficient of expansion [43]. Under the biaxial strain, the band gap of $\alpha$-$Si_2N$ approximately decreases linearly in the range of 0%-11%, and 11%-16%, respectively, as shown in Fig. 7. Similarly, the band gap of $\beta$-$Si_2N_2$ also decreases approximately linearly with a breakpoint at 7%. Both band gaps close at 16%, where insulator-metal transitions occur. The decrease of gap may be caused by the change from $sp^3$ hybridization to $sp^2$ hybridization. As strain increases, the monolayer gradually becomes two $sp^2$ hybridized monolayers similar to graphene, thus the band gap decreases. From the total charge density at 16% tensile strain, as shown in Fig. 8, more charges are localized on the N, and less charges are localized on the bonding region of the Si-Si bond, which indicates that the Si-Si bond is weakened. The transition strain is much less than the ultimate strain of graphene, which is up to 25% [44]. The proposed materials are bilayers with a buckled structure, which may help the materials to withstand a larger strain. If the materials are attached to substrates, they may be stabilized by the substrate, and possibly achieve high strain without fracture.

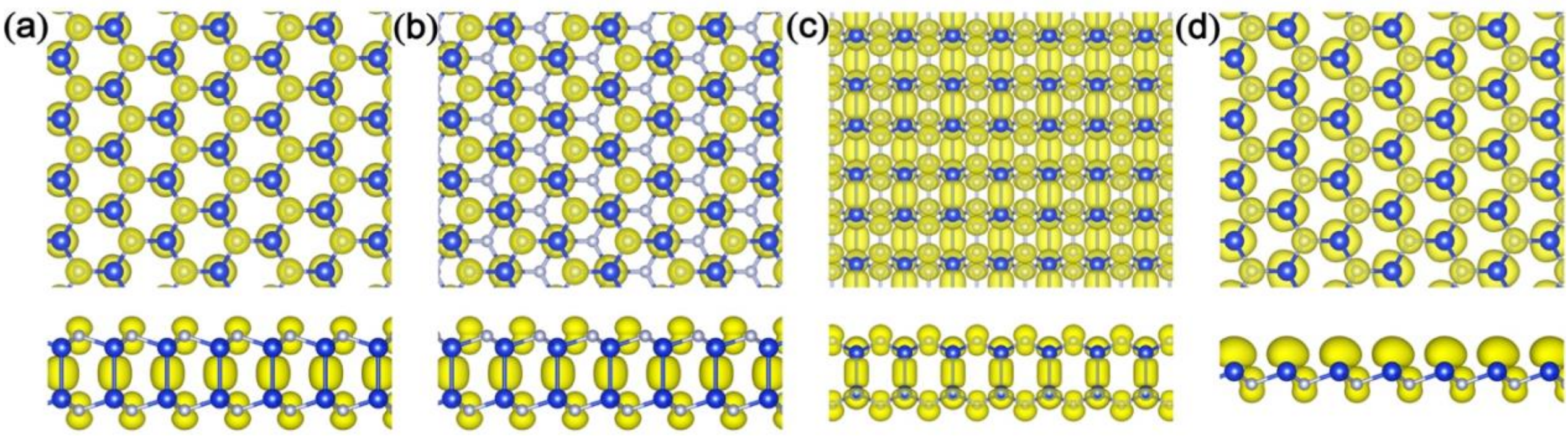

Figure 6. The band-decomposed charge density in the energy window of -2 eV to 0 eV with respect to the Fermi energy of (a) α-$Si_2N_2$, (b) β-$Si_2N_2$, (c) γ-$Si_4N_4$, and (d) SiN monolayer. The upper and lower panels denote the top view and side view of the charge density, respectively. The gray and blue spheres represent the nitrogen and silicon atoms, respectively.

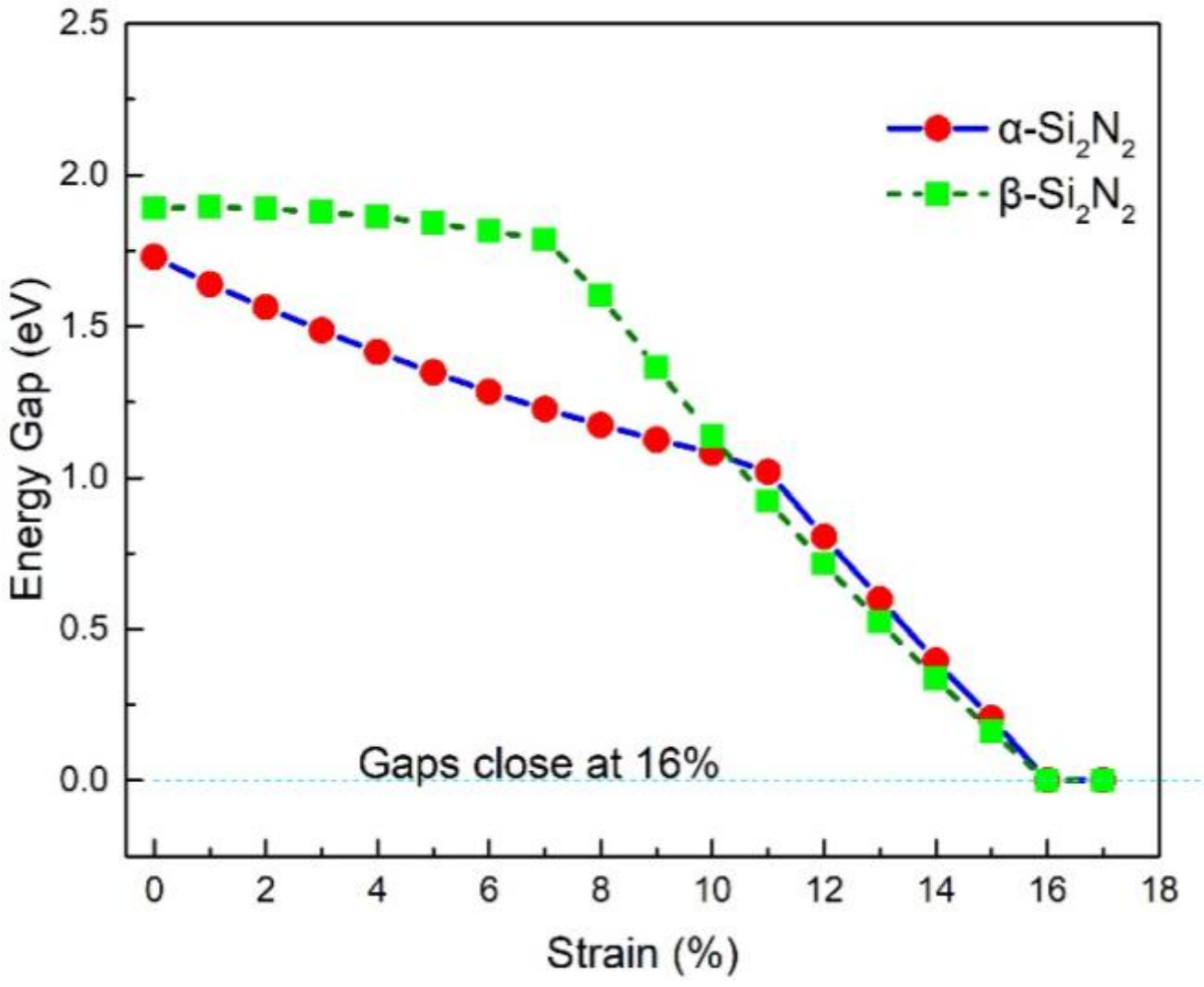

Figure 7. The band gaps depend on the biaxial tensile strain. The red dot and green block denote the band gaps evolution of α-$Si_2N_2$ and β-$Si_2N_2$, respectively.

## 4. Conclusion

This work has proposed three 2D silicon nitrides, α-$Si_2N_2$, β-$Si_2N_2$, and γ-$Si_4N_4$, based on the density functional theory. The phonon dispersion indicates that the α-$Si_2N_2$ and β-$Si_2N_2$ are dynamically stable, while γ-$Si_4N_4$ may exist for a long time while attaching to the substrates. Both α-$Si_2N_2$ and β-$Si_2N_2$ are equally stable, while the conversion between them is difficult because of a high energy barrier. The sound velocities of the acoustic modes indicate that both α-$Si_2N_2$ and β-$Si_2N_2$ are isotropic and have strong rigidities comparable to graphene. The SiN monolayer with a spin-polarized band structure shows a sub-lattice magnetization of $1\ \mu_B$, which arises from the dangling bonds on silicon atoms, but may be highly unstable. As two SiN monolayers are connected to form either α-$Si_2N_2$ or β-$Si_2N_2$, it becomes spin-degenerate. Both α-$Si_2N_2$ and β-$Si_2N_2$ have intrinsic indirect band gaps of 1.73 eV and 1.89 eV, respectively. Furthermore, both α-$Si_2N_2$ and β-$Si_2N_2$ can undergo insulator-metal transitions under the sufficient tensile strain. These bilayers may apply in spintronics and straintronics due to their novel lattice structures and electronic properties. Similar works of α-$Si_2N_2$ [45, 46] also reported after our work announced (arXiv:1707.02819).

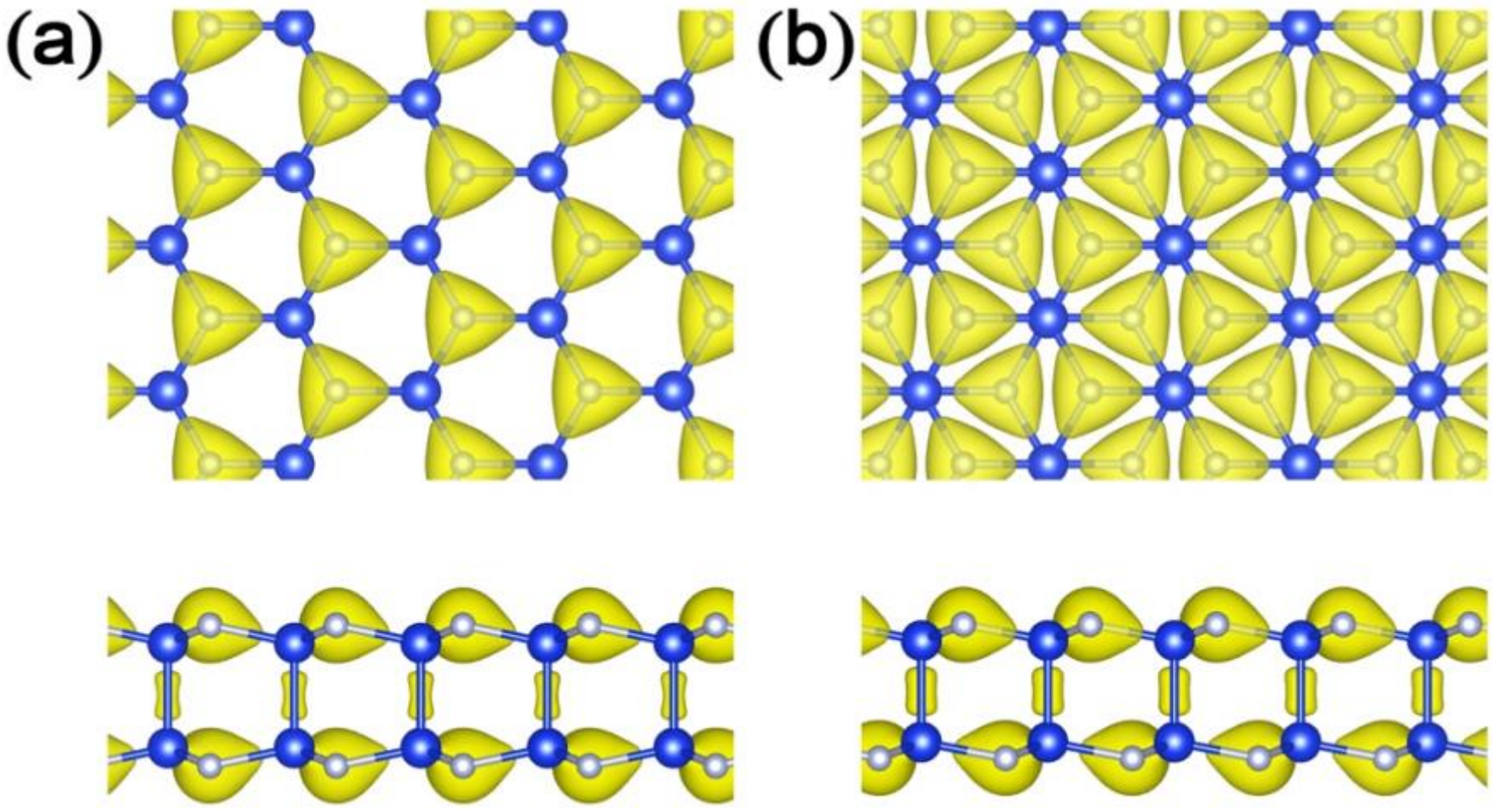

Figure 8. The total charge density of (a) α-$Si_2N_2$ and (b) β-$Si_2N_2$ under 16% tensile strain, respectively. The upper and lower panels denote the top view and side view of the charge density, respectively. The gray and blue spheres represent the nitrogen and silicon atoms, respectively.

**CRediT authorship contribution statement**

**Jiesen Li:** Conceptualization, Methodology, Software, Validation, Formal analysis, Investigation, Resources, Data Curation, Writing - Original Draft, Writing - Review & Editing, Visualization. **Wanxing Lin:** Conceptualization, Methodology, Validation, Formal analysis, Investigation, Data Curation, Writing - Original Draft, Writing - Review & Editing, Visualization, Funding acquisition. **Junjun Shi:** Formal analysis. **Feng Zhu:** Formal analysis. **Haiwen Xie:** Formal analysis, Investigation. **Dao-Xin Yao:** Conceptualization, Methodology, Formal analysis, Resources, Writing - Original Draft, Writing - Review & Editing, Project administration, Funding acquisition

## Declaration of competing interest

The authors declare that they have no known competing financial interests or personal relationships that could have appeared to influence the work reported in this paper.

## Acknowledgments

We thank Shi-Dong Liang and Bao-Tian Wang for helpful discussions. This work is supported by NKRDPC-2022YFA1402802, NSFC-92165204, NSFC-12304079, Shenzhen International Quantum Academy (Grant No. SIQA202102), and Leading Talent Program of Guangdong Special Projects (No. 201626003). Wanxing Lin is also supported by program for scientific research start-up funds of Guangdong Ocean University. Most of the calculations were performed at Tianhe-II in Guangzhou.

Supplementary Material for

# Phonon and electronic properties of semiconducting silicon nitride bilayers

Jiesen Li [1,#], Wanxing Lin [2,3,#], Junjun Shi[1], Feng Zhu[1], Haiwen Xie[1], Dao-Xin Yao [2,4 †],

[1] School of Environment and Chemical Engineering, Foshan University, Foshan 528000, P. R. China

[2] State Key Laboratory of Optoelectronic Materials and Technologies, Guangdong Provincial Key Laboratory of Magnetoelectric Physics and Devices, School of Physics, Sun Yat-Sen University, Guangzhou 510275, P. R. China

[3] School of Materials Science and Engineering, Guangdong Ocean University, Yangjiang 529500, P. R. China.

[4] International Quantum Academy, Shenzhen 518048, P. R. China.

[#] Two authors contribute equally.

[†] Corresponding author. Tel: 86 020 84112078. E-mail: yaodaox@mail.sysu.edu.cn (Dao-Xin Yao)

## I. AIMD simulations

To verify the structural stabilities, *ab initio* molecular dynamics (AIMD) simulations at 300K are performed for the 4×4×1 supercell using the algorithm of Anderson thermostat starting with 300K, and the simulation lasted for 4ps with a time step size of 1 fs.

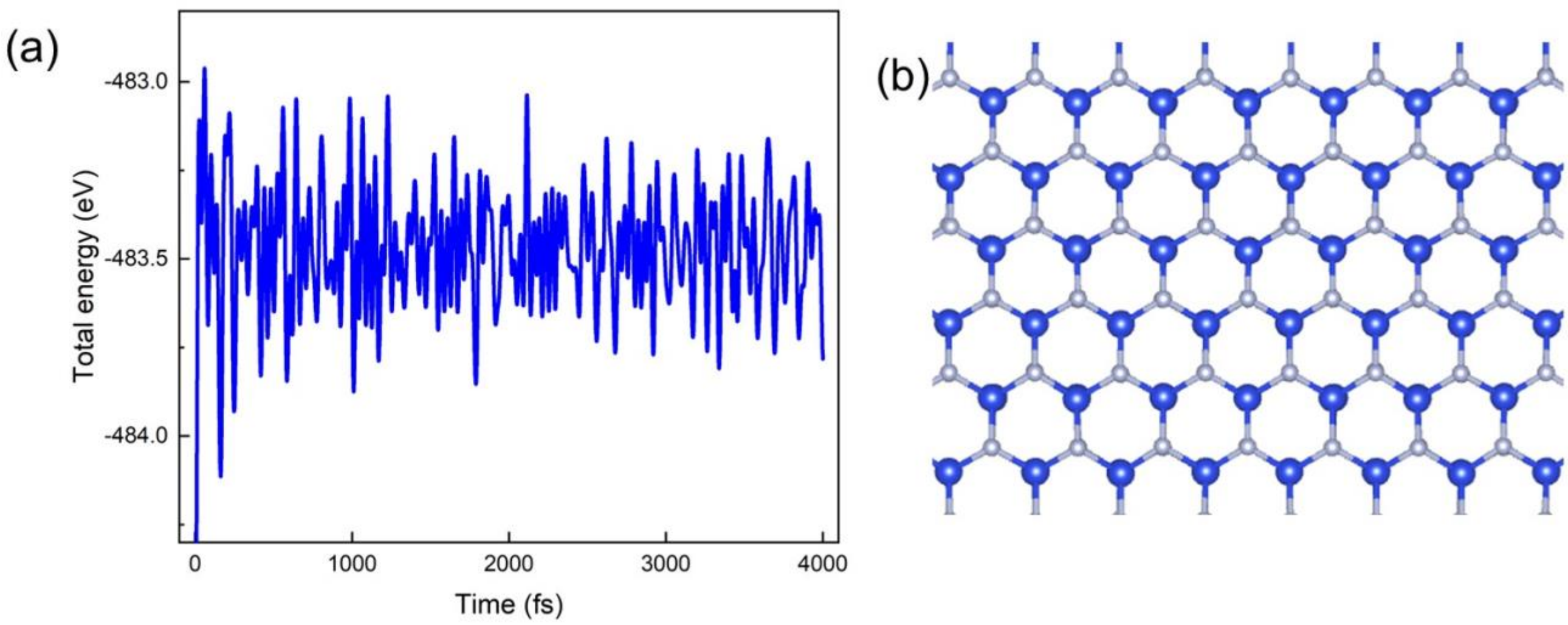

Figure S1. (a) The variation of the total energy of α-$Si_2N_2$ with time in AIMD simulations at 300 K, (b) shows the top view of the final geometry structure.

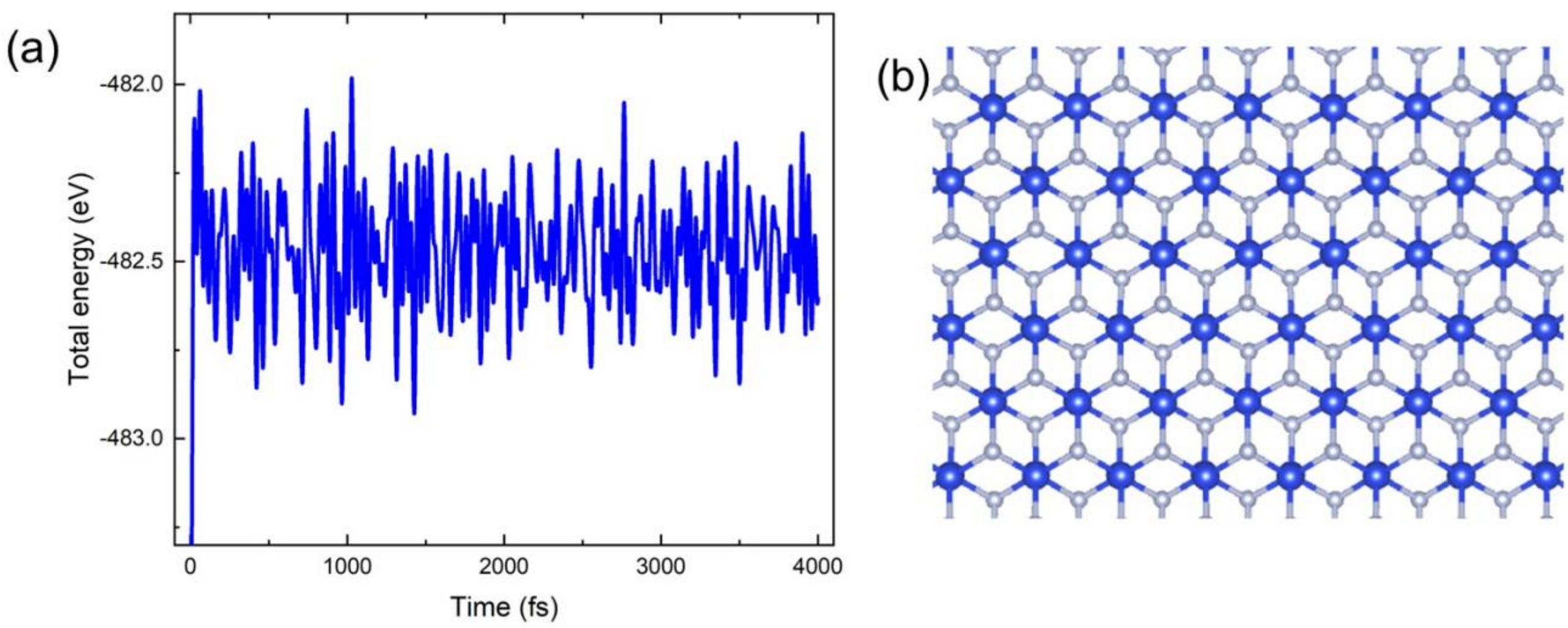

Figure S2. (a) The variation of the total energy of β-$Si_2N_2$ with time in AIMD simulations at 300 K, (b) shows the top view of the final geometry structure.

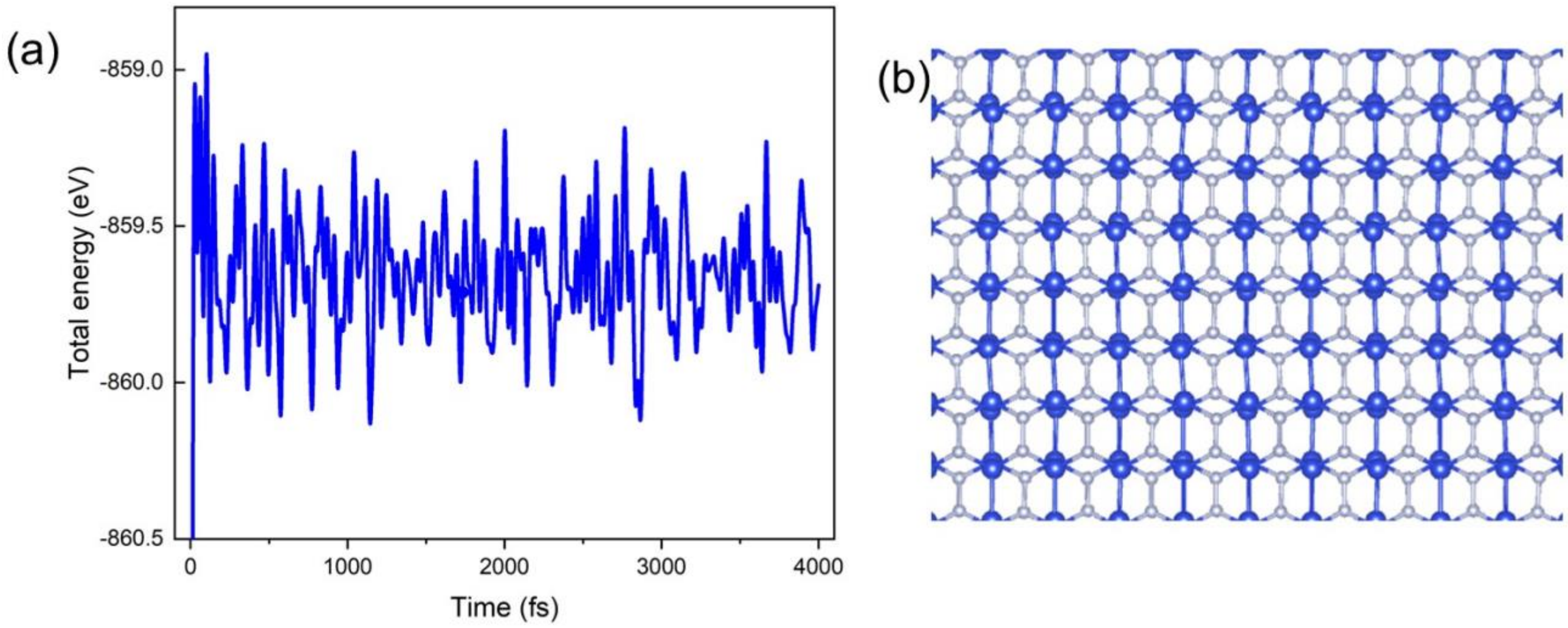

Figure S3. (a) The variation of the total energy of γ-$Si_4N_4$ with time in AIMD simulations at 300 K, (b) shows the top view of the final geometry structure.

II. Projected density of states

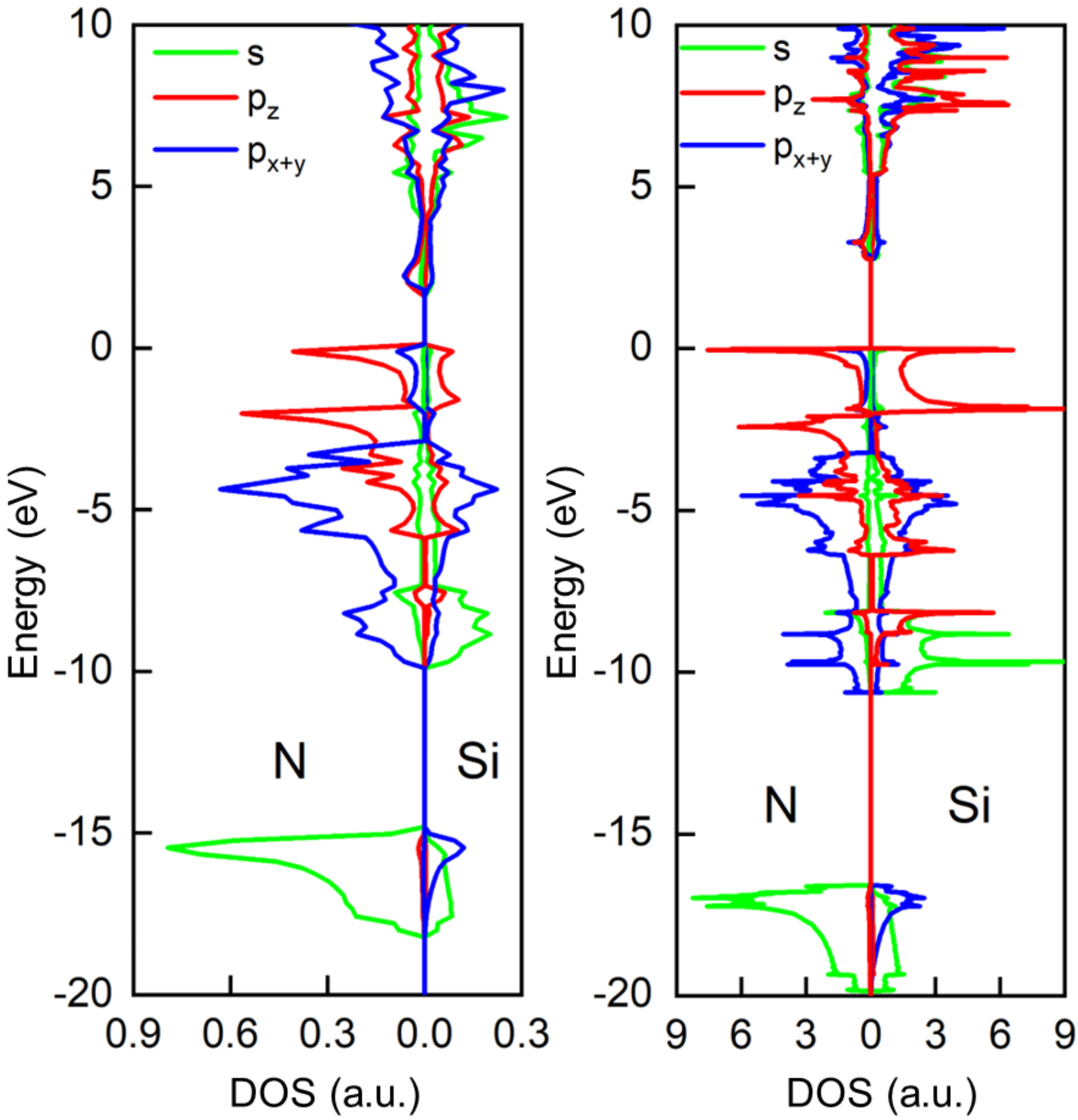

Figure S4. The PDOS of α-$Si_2N_2$ on the PBE (left) level and HSE (right) level, respectively.

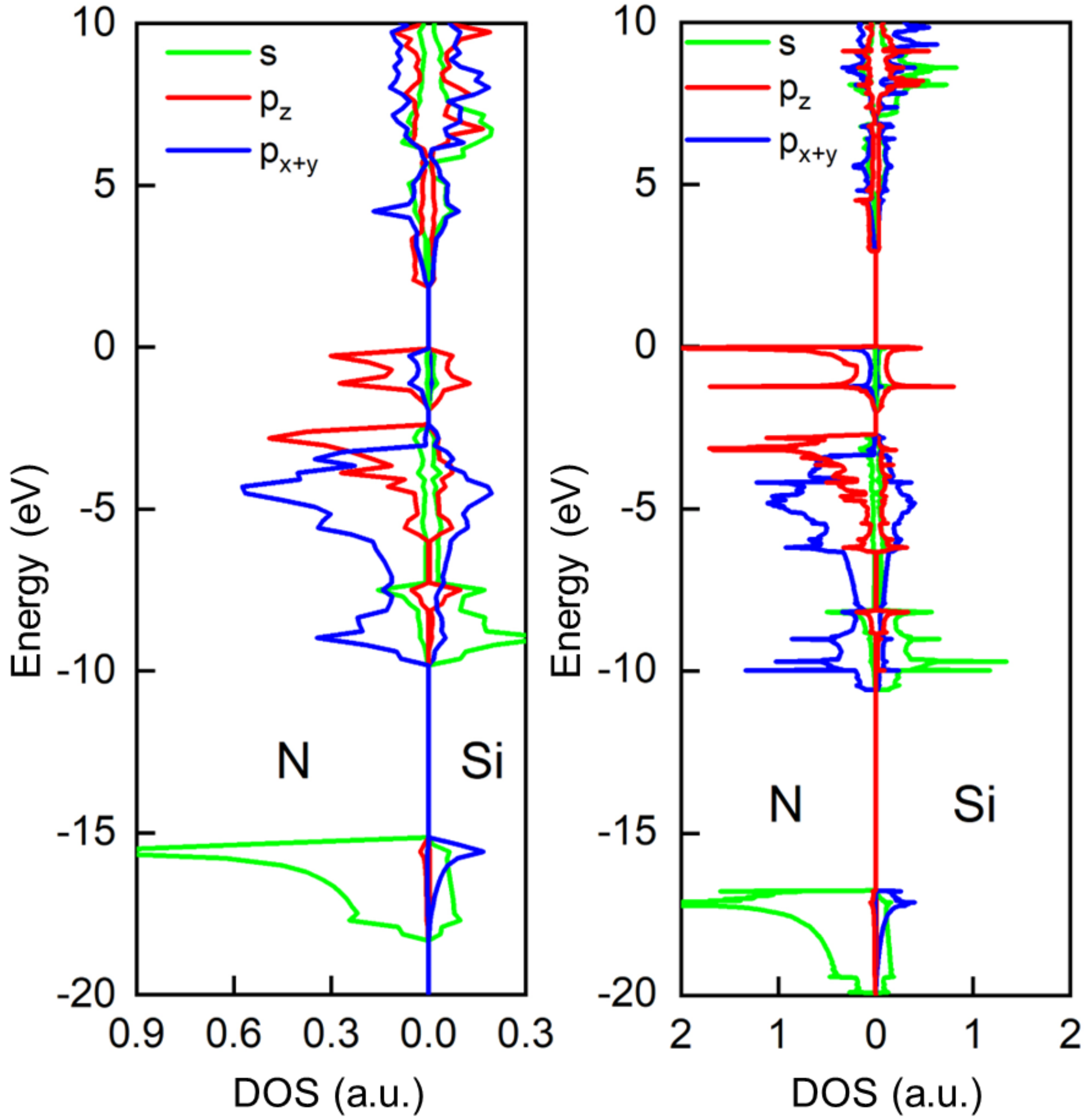

Figure S5. The PDOS of β-$Si_2N_2$ on the PBE (left) level and HSE (right) level, respectively.

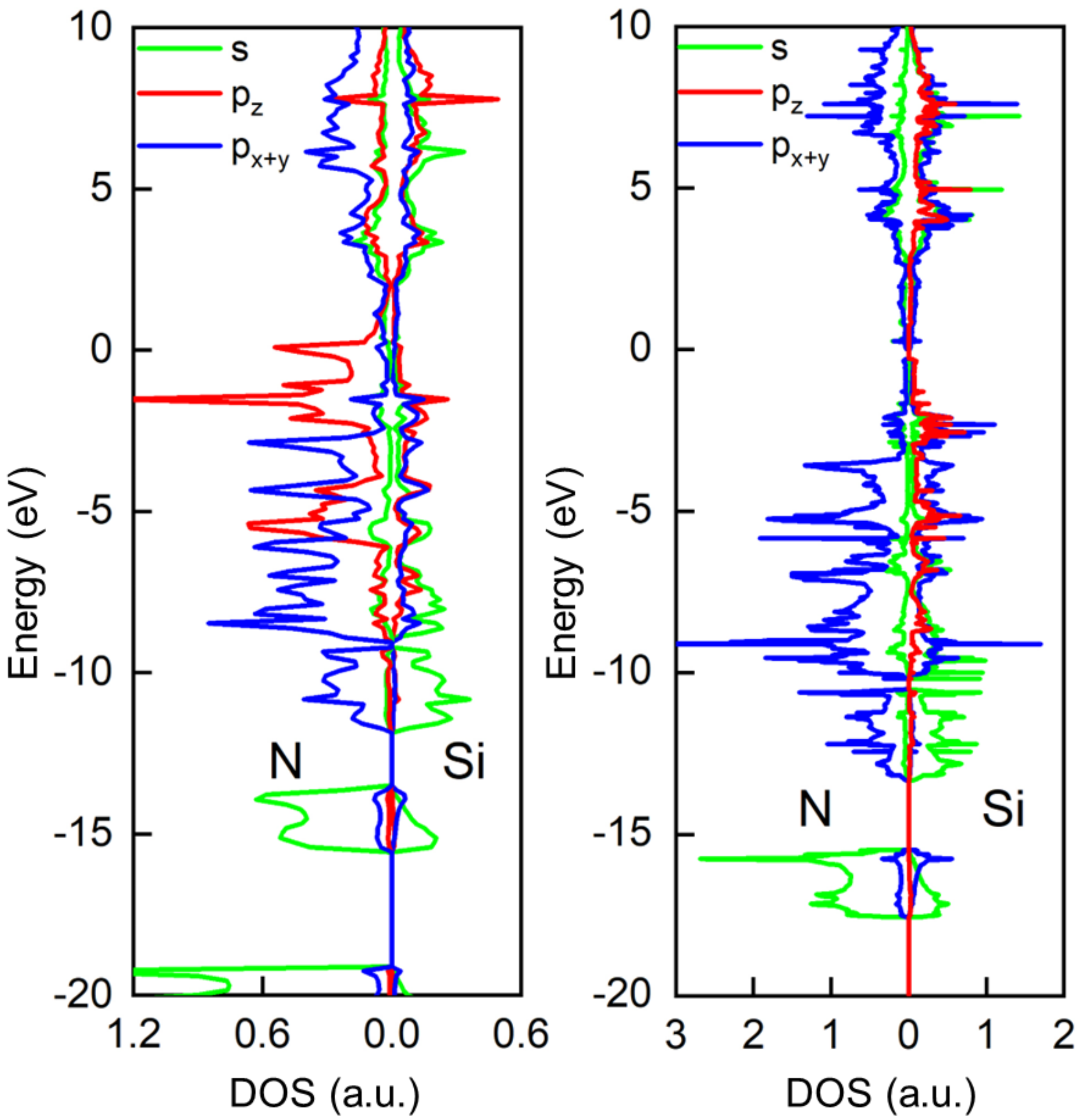

Figure S6. The PDOS of γ-$Si_4N_4$ on the PBE (left) level and HSE (right) level, respectively.

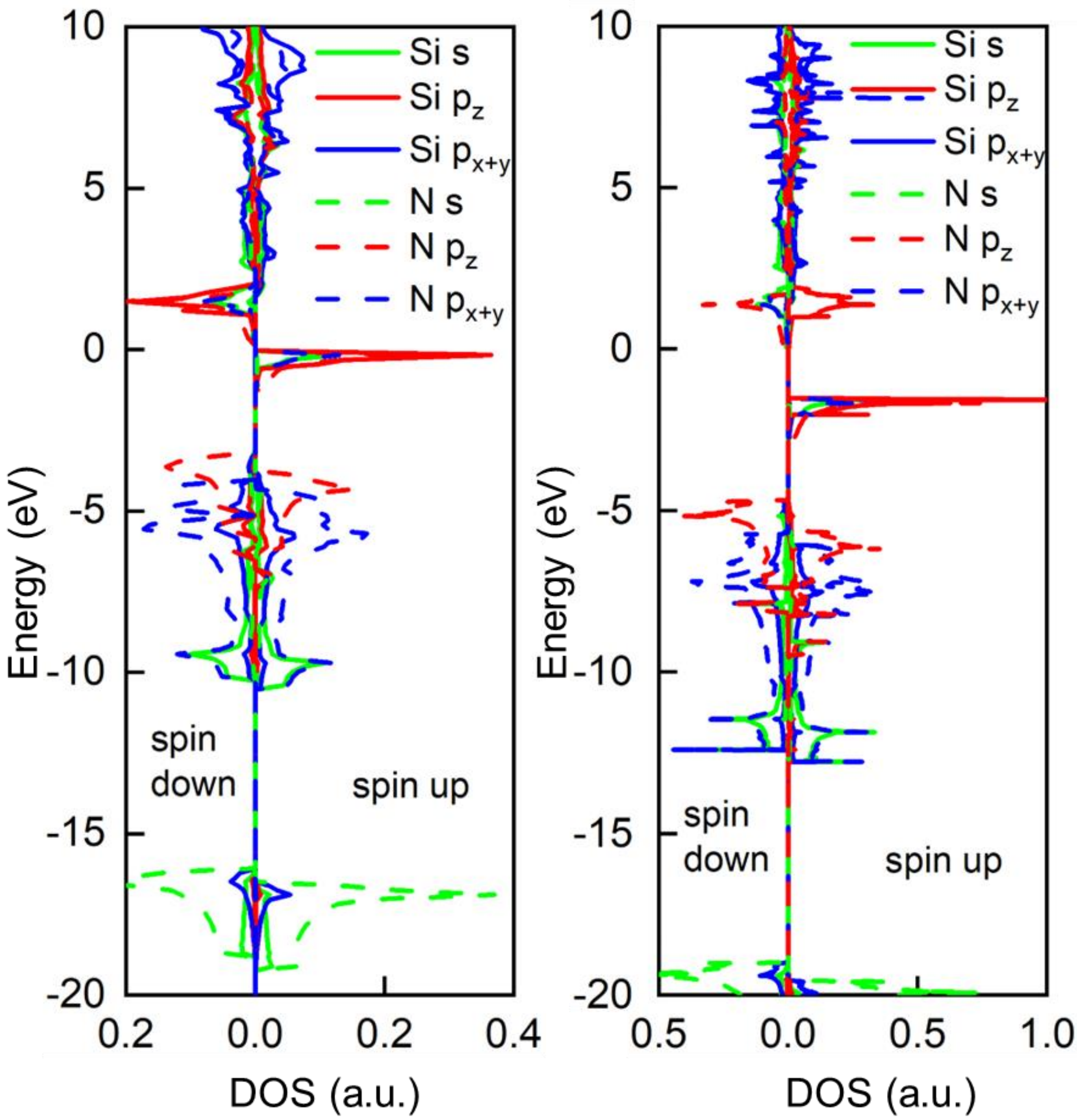

Figure S7. The PDOS of SiN monolayer on the PBE (left) level and HSE (right) level, respectively.